# In-plane nanoelectromechanical resonators based on silicon nanowire piezoresistive detection

E Mile*, G Jourdan*, I Bargatin *[+1], S Labarthe *, C Marcoux *, P Andreucci *, S Hentz*, C Kharrat*, E Colinet*, and L Duraffourg*

*CEA/LETI MINATEC 17 rue des Martyrs. 38054 GRENOBLE CEDEX 9.
[+]Condensed Matter Physics 114-36, California Institute of Technology, Pasadena, California 91125

**Email**. laurent.duraffourg@cea.fr

**Abstract.** We report an actuation/detection scheme with a top-down nano-electromechanical system for frequency shift-based sensing applications with outstanding performance. It relies on electrostatic actuation and piezoresistive nanowire gauges for in-plane motion transduction. The process fabrication is fully CMOS compatible. The results show a very large dynamic range (DR) of more than 100dB and an unprecedented signal to background ratio (SBR) of 69dB providing an improvement of two orders of magnitude in the detection efficiency presented in the state of the art in NEMS field. Such a dynamic range results from both negligible 1/f-noise and very low Johnson noise compared to the thermomechanical noise. This simple low-power detection scheme paves the way for new class of robust mass resonant sensor.

**Key words.** Nano-electromechanical system, resonator, silicon nanowire

**PACS.** 62.23-c , 63.25-g

---

[1] Present address: Stanford University – Stanford California 94305-4070

## 1. Introduction

NEMS are actively being explored due to their incredible potential for applications such as ultrasensitive mass [1], [2], [3], [4] and force sensing [5]. However, efficient actuation and sensitive detection at the nanoscale remains a challenge. The small displacements of these miniaturized devices induce very low signals which are overwhelmed by parasitic background. A lot of effort has been devoted to develop new transduction and background reduction [6]. A variety of NEMS detection techniques, such as capacitive [3], [7] [8], magnetomotive [9], piezoresistive [10], [11] and field-emission [4] [12] transduction, have been proposed. Magnetomotive typically requires large magnetic fields (2-8 T) and is thus not suitable for integrated applications. The field-emission effect detection demands complex instrumentation and its stability in time is still questionable. Moreover, this technique uses bottom-up approach that is hardly compatible with large scale integration (VLSI) process. Piezoresistive detection scheme offers great potential compared to capacitive one especially at high resonant frequency measurements [10] [13].

Recently, mass resolution down to $7zg/\sqrt{Hz}$ [1] has been demonstrated using a metallic gauge layer deposited on the top of a cantilever. Another approach [14] consists in using a doped silicon nanowire that produces a second-order piezoresistive effect for large displacements of the nanowire. However to date bottom-up nanowire cannot be fabricated using a VLSI process compatible with a standard CMOS technology.

In this paper, we demonstrate an original technique of highly efficient in-plane motion detection based on suspended p++ doped piezoresistive nanowires connected in a symmetric bridge configuration to a resonating lever arm. The differential bridge architecture provides intrinsic signal amplification and background suppression. We show that detection through silicon gauges has a better signal to noise ratio at room temperature than the metallic layer used as piezoresistive gauge. Although the Johnson noise is higher with semiconductor nanowire gauges (due to their larger resistance) the increase in signal is much larger than the increase in noise. We therefore present an alternative way using piezoresistive technique showing similar performance as metallic gauges. We therefore reconsider the belief that metallic gauges are the best candidates for nanoscale piezoresistive transduction.

In addition, in-plane motion architecture offers more flexibility of design and simplifies process development. Our device uses CMOS-based fabrication and is therefore fully compatible with very large scale integration (VLSI) of NEMS on 200mm wafer in the future.

This paper starts with an overview of the fabrication process and architecture, continues with measurements and results, and concludes with a discussion of the efficiency of the detection scheme and the frequency stability of these devices.

**2. NEMS resonator and principle of operation**

*2.1. Device*

Advances in top-down lithographic processes have enabled fabrication of nanostructures with sizes similar to those achieved with bottom-up synthesis methods. The NEMS device presented in this paper is fabricated using CMOS compatible materials with nano-electronics state-of-the-art lithography and etching techniques. We used a 200-mm silicon-on-insulator (SOI) wafer of (100) orientation with a 160-nm-thick top silicon structural layer (resistivity ≈ 10 Ω·cm) and a 400-nm-thick sacrificial oxide layer. The top silicon layer was implanted with boron ions (p-type) through a thin layer of thermal oxide. Homogenous doping (~$10^{19}$cm$^{-3}$) in the whole thickness of the top silicon was obtained through specific annealing step (for material reconstruction and doping activation commonly used in CMOS technology), resulting in top layer resistivity of approximately 9 mΩ·cm. A hybrid e-beam/DUV lithography technique (allowing 50 nm minimum feature size) was used to define the nano-resonators and electrode pads, respectively. Top silicon layer was etched by anisotropic reactive ion etching (RIE). In order to decrease the lead resistances, the interconnecting leads have been made thicker with a 650 nm thick AlSi layer, a typical metal for CMOS interconnections process. Finally, the nanoresonators have been released using a vapor HF isotropic etching to remove the sacrificial layer oxide beneath the structures.

1500 devices per wafer of such design are fabricated with this VLSI process. Functionality of the final devices is checked measuring both the lead and gauges resistances and resonant frequency. The yield is 95% per wafer in average.

The NEMS is composed of a fixed-free lever beam and two piezoresistive gauges connected to the cantilever at a distance $l_1=0.15l$ from its fixed end where $l$ is the beam length (see TAB. 1). This value was chosen to maximize the stress inside the gauges due to the cantilever motion (see FIG. 1). The gauges have been etched along the <110> direction in order to benefit from the high gauge factor associated with $p^{++}$ doped silicon. A drive electrode was patterned along one side of the vibrating beam for electrostatic actuation. The general architecture is given in FIG. 1 and the device dimensions are summed up in TAB. 1.

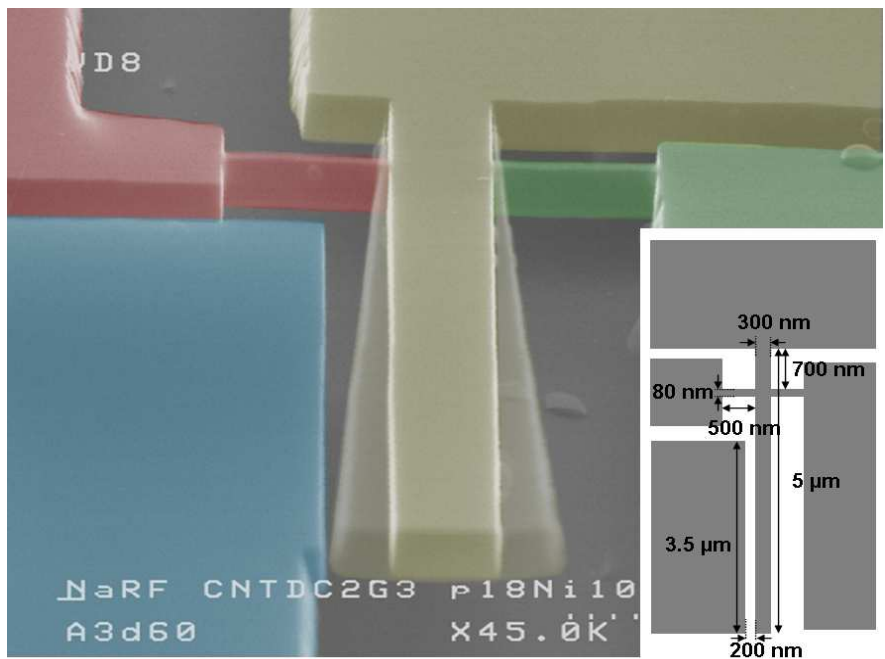

**FIG.1** – Artificially colored and modified SEM image illustrating the in-plane vibration of the beam

**TAB. 1** – Typical values of the device

| Beam length $l$ | Beam width $w$ | Distance Anchor/Gauges $l_1$ | Gauge length $b$ | Gauge width $w_1$ | Electrode length $a$ | Gap Electrode/Beam $g$ |
|---|---|---|---|---|---|---|
| 5μm | 300nm | 700nm | 500nm | 80 nm | 3.5 μm | 200nm |

The lead resistance of approximately 4 kΩ and the gauge resistance of 3.6 kΩ were measured using the 3-point local AFM technique [15].

*2.2. Principle of operation*

To evaluate the dynamical behavior of the NEMS geometry shown in, we used a model based on Euler-Bernoulli beam theory that is detailed in Ref [13]. From this model, we can easily compute the first Eigen frequency, $\omega_0$, as well as the force $F_g$ acting on the gauges,

$$F_g(\omega) = \alpha \frac{\omega_0^2}{\omega_0^2 - \omega^2 + j\omega\omega_0/Q} F_{el}(\omega) \tag{1}$$

where $\alpha$, $\omega$, $Q$ and $F_{el}(\omega)$ are the amplification factor, the angular frequency (rad/s), the quality factor, and the electrostatic driving force respectively.

The electrostatic driving force along the lever beam is given by $F_{el} = 1/2\, C'V^2$, where $V$ the applied voltage and $C'$ is the derivative of the capacitance $C$ between the cantilever and the drive electrode with respect to the lateral displacement. At resonance, $\omega = \omega_0$ and the force amplification is given by $F_g(\omega_0)/F_{el}(\omega_0) = j\alpha Q$ \hfill (2)

A comparison with the results of Finite Element Modeling (FEM) validated to a large extent our analytical model, as shown in TAB 2.

**TAB. 2** – Comparison of predictions of analytical and FEM models - $M_{eff}$ is the effective mass

|  | $\omega_0/2\pi$ | $\alpha$ | $M_{eff}$ |
|---|---|---|---|
| Analytical model | 21.10 MHz | 6.05 | 140 fg |
| FEM model | 20.65 MHz | 5.2 | NA |

The slight discrepancies are due to the assumption that there is no bending moment introduced by the gauges with a perfect anchor.

This design results in first order piezoresistive effect (as opposed to weaker second order like in [9]) with the suspended gauges acting as collectors of the stress $F_g/s$, where $s$ is the cross section area of the gauges. The strain induced in the gauges is transduced into a resistance variation $\Delta R$ through the piezoresistive effect:

$$\frac{\Delta R(\omega)}{R} = \gamma \cdot \varepsilon(\omega) = \gamma \cdot \frac{F_g(\omega)}{2 \cdot s \cdot E} \tag{3}$$

where $\gamma$ and $E$ are the gauge factor and the Young's modulus of the gauges, respectively. The piezoresistive factor $\gamma$ is usually written as,

$$\gamma = (1+\nu) + \frac{1}{\varepsilon}\frac{\Delta\rho}{\rho} \qquad (4)$$

where $\rho$, $\varepsilon$ and $\nu$ are the resistivity, the strain and the Poisson ratio respectively. The gauge factor, which links the mechanical strain in a gauge to its relative resistance change, is caused by two effects. The first is a purely geometric effect and is associated with elastic deformation (first term in parenthesis in eqn 4), while the second corresponds to the modification of the energy bands inside the semiconductor, which alters its resistivity (second term in eqn 4). In metals, only the first term is significant, and the gauge factor ranges from 1 to 4. In semiconductors, the second term is the most significant contribution. For the chosen <110> crystalline orientation and the doping level of $10^{19}$ cm$^{-3}$, the theoretical value is 47 [16]. In our case, $\gamma$ is evaluated to be around 40 from the amplitude peak at the resonance using equations (1) and (2). This experimental result is in good agreement with the theory. Values of material parameters used in this article are summed up in TAB 3.

TAB. 3 – Parameters of the cross-beam NEMS

| Parameters | $E$ (GPa) | $\nu$ | $\mu$ (g.cm$^{-3}$) | $\rho$ (mΩ.cm) | $\gamma$ |
|---|---|---|---|---|---|
| | 169 | 0.26 | 2330 | 1.4 | 40 |

The device under test was connected to a radio frequency (RF) circuit board through wire bonding and loaded into an RF vacuum chamber for room temperature measurements. At high frequencies, the electrical readout is complicated by parasitic capacitances which change the expected behavior of the electrical circuit. Given the cable capacitance (100pF/m), the input impedance of the Stanford Research 830 lock-in amplifier ($R$=10MΩ, $C$=25pF), and the device pads, the overall parasitic capacitance at the NEMS output is close to $C_p$=125pF. This capacitance combines with the electrical resistance of the setup to produce a low pass-filter on the output signal with a cut-off frequency of 120 kHz. To avoid parasitic impedances and cross talk, we used a 2ω down-mixing technique to read out the resistance variation at a lower frequency $\Delta\omega$ (typically between 10 kHz and 30 kHz) [17]. A schematic of the setup is shown in Fig. 2. The cross beam is actuated with a drive voltage $V_d$ at ω/2. Because the electrostatic force is proportional to $V_d^2$, the strain in the gauges varies at the frequency ω. This technique results in efficient frequency decoupling of the downmixed signal from parasitic feedthrough. The downmixed signal read out at the middle of the bridge is proportional to,

$$V_{out}(\Delta\omega) \propto \Delta R \cos(\omega t) \cdot I_b \cos((\omega + \Delta\omega)t) = \frac{1}{2} I_b \cdot \Delta R \cdot \cos(\Delta\omega t) \qquad (5)$$

where $I_b$ is the bias current through the gauges induced by the bias voltage $V_{bias}$ (see Fig. 2).

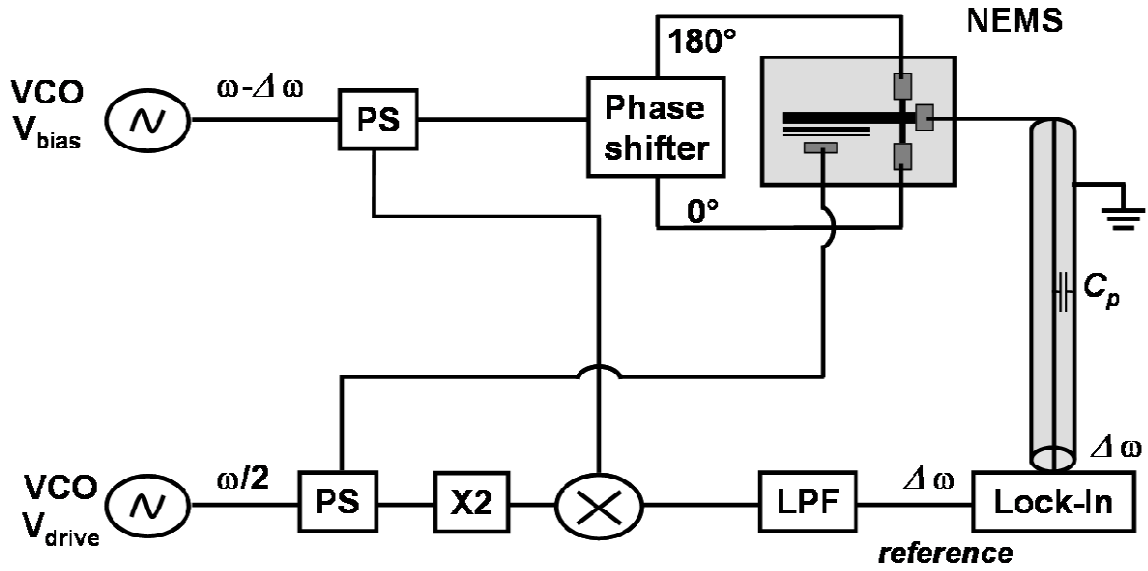

**FIG. 2** – Schematic of the experimental setup used for detecting the resonant motion of the NEMS in Fig 1. PS, LPF, VCO, X2 are power splitter and phase shifter, voltage control oscillator and frequency doubler respectively.

The two gauges on each side of the lever arm work under equal and opposite tensile and compressive strains. This balanced bridge configuration suppresses the parasitic feedthrough at the middle point of the bridge.

### 3. Experimental results

The experiments, performed at room temperature and pressure of less than 1 mTorr, showed a remarkably small and flat background, as shown in FIG. 3. The measured quality factor was approximately 5000 in vacuum and 200 at atmospheric pressure. Quality factors up to 10000 were measured.

*3.1. Signal to background ratio*

The geometrical and frequency decoupling between the actuation and detection results in a very large signal to background ratio (SBR) of 67dB. For ultra low mass sensing, *SBR* is an important parameter that should be maximized. At the resonance, a large *SBR* means large variation of the phase for a small frequency shift (Bode representation). In a closed loop (phase locked loop for instance) the digital error on the readout of the phase will be then low with devices having a large *SBR*. Furthermore device with large background will be more sensitive to the random perturbations of its environment. This value is close to two orders of magnitude larger than previous *SBR* [3] [18] [19] at ambient temperature (300K). Average value per wafer of resonant frequency is 19.16 MHz with a maximum dispersion of 2% showing the pretty good reproducibly of the VLSI process.

$V_{drive}$ can be set between a few hundred millivolts to 5 volts before having non linear behavior of the cantilever. $V_{bias}$ can be set up to 10 V before gauge melting. In the experiment, the voltages are set to a value of 1.5V, which corresponds to the maximum supplied by our AC-generator.

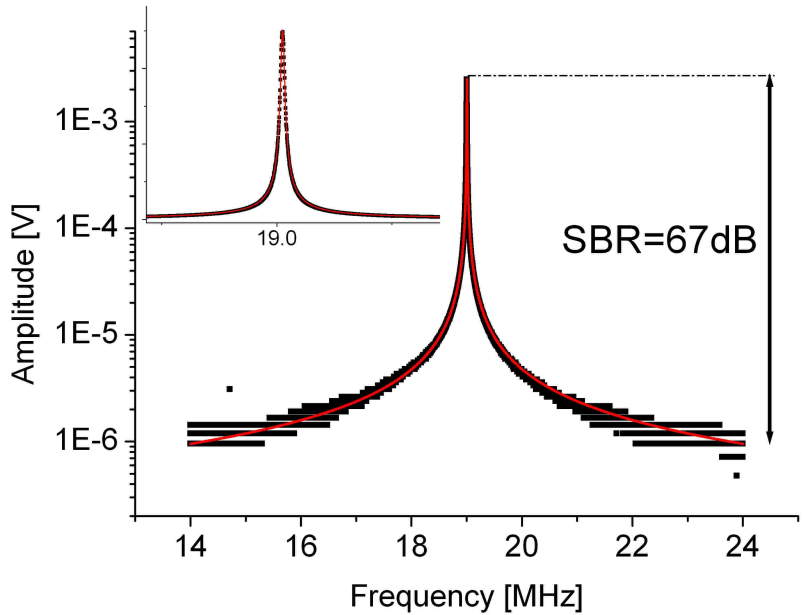

**FIG. 3** – Typical output signal from the structure shown in Fig. 1 in a vacuum with pressure under 1 mTorr. The signal to background ratio is 67 dB for $V_{drive}$=1.5V and $V_{bias}$=1.5V – Sampling time=30 ms. The inset shows the same data using linear scale.

*3.2. Noise and signal to noise ratio*

For frequency-shift based sensing applications, frequency fluctuations naturally impose a limit on the sensitivity. One source of frequency fluctuations is due to a finite signal-to-noise ratio (SNR) at resonance and the resolution can be defined with the approach presented in Ref. [5]. As shown in Fig. 4, a large SNR of around 100dB can be obtained with our device. This value is larger than data reported previously (see [1], [10] for example). To measure the noise, we followed the technique described in Ref. [17]. There was no external drive, and only a bias voltage was applied to the gauges. The noise $(V/\sqrt{Hz})$ was then measured by sweeping the frequency of the bias signal $\omega_{bias}$ while keeping a constant offset frequency of $\Delta\omega/2\pi = 25 kHz$. As a result, the high-frequency thermo-mechanical noise was mixed down to a lower frequency $|\omega_{bias} - \omega_0| = \Delta\omega$. We thus obtained two peaks with amplitudes of 28 nV/√Hz, separated by $50 kHz$ (see inset of Fig. 4). The noise level is evaluated over a 1 Hz-bandwidth.

The noise floor $S_d^{1/2} \approx 13 nV/\sqrt{Hz}$ resulted from both the Johnson noise and input noise of detection electronics. The thermomechanical noise $S_{th}^{1/2}$ can be calculated from the peak amplitude and the floor level and is approximately 24.8 *nV/√Hz*. The Johnson noise is given by $S_J^{1/2} = \sqrt{4k_B TR}$ $\approx 11.2 nV/\sqrt{Hz}$ (R~7600Ω). The electronics noise is then $S_V^{1/2} = \sqrt{S_d - S_J} \approx 6 nV/\sqrt{Hz}$, which agrees with the noise level specified by the manufacturer of the lock-in amplifier.

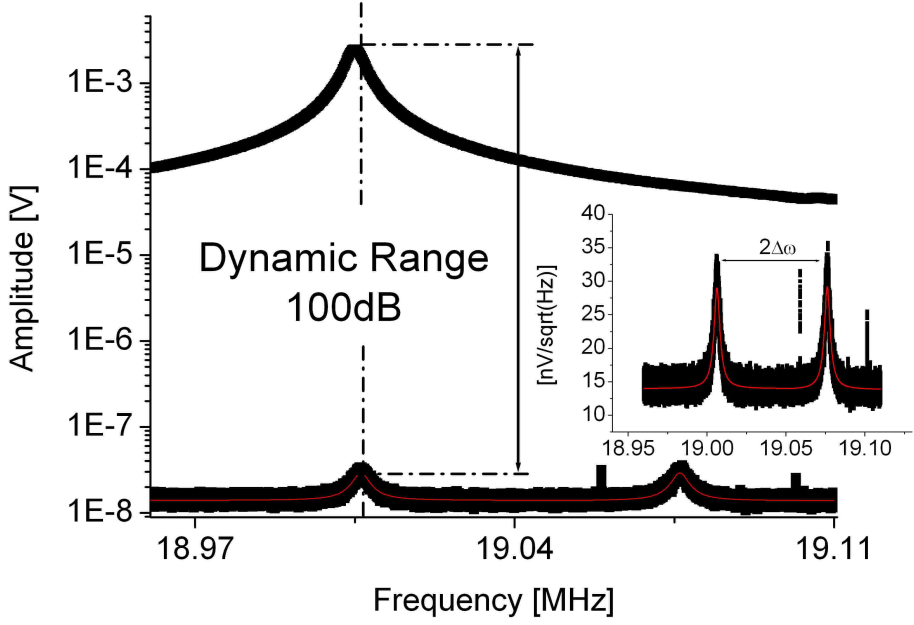

**FIG. 4** – Signal to noise ratio obtained for $V_{drive}$=1.5V and $V_{bias}$=1.5V – Noise is computed for 1Hz-bandwith – The inset corresponds to the noise density peaks around the resonance frequency.

Typically, 1/f-noise created by resistance fluctuations is the main limitation in piezoresistive sensors [9]. However, these resistance fluctuations were not observed in our devices at 20-MHz operating frequency. In order to investigate the consistency of such a result, we computed 1/f-noise density using Hooge's empirical relation [20],

$$S_H = \frac{HV_{bias}^2}{N|f - f_{bias}|} \quad (6)$$

where $N$ is the total number of carriers within the gauge and $f_{bias}$ is the bias frequency. The Hooge parameter $H$ is extracted from the measurement of the relative resistance variation according to the readout voltage frequency for two amplitudes (see Fig. 5). An AC-bias (~15 kHz) is used to remove the 1/f-noise of the lock-in. By linearly fitting the data, we find $H$ to be approximately $10^{-6}$. From Eqn. (6), we then estimate the resulting noise to be a few nV/√Hz at 20 MHz, which is negligible compared to other source of noise. To illustrate this, we included the noise floor level (Johnson and electronics noises) and the thermomechanical noise level in terms of relative resistance fluctuation in Fig. 5. For frequencies higher than 100 kHz 1/f-noise appears to be lower than other noises. This result is in particular obtained thanks to homogenous doping ($10^{19}$cm$^{-3}$) in the whole thickness and specific annealing. Bad doping process in conjunction with low doping level could lead to opposite conclusion.

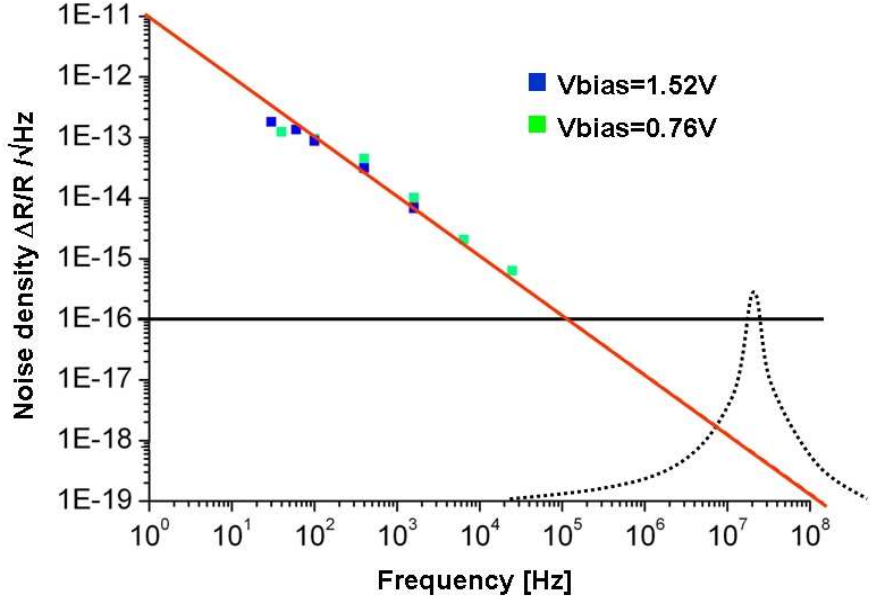

**FIG. 5** – Contribution of different noise sources expressed in relative resistance change, which is independent of $V_{bias}$. *1/f* noise density measurement for different bias voltages (colored squares) compared with both the noise floor and the thermomechanical noise. Red curve is the linear fit of the experimental data for *1/f* noise. Black curve corresponds to the noise floor (i.e. electronic and Johnson noises). Black dashed curve corresponds to a schematic of the thermomechanical noise.

It is important to note that we obtain a priori an unexpectedly large SNR (see Fig.4). For our semiconductor nanowire gauges, we infer the piezoresistive gauge factor $\gamma$ to be approximately 40, compared to at most a few unities for metallic-layer piezoresistors. The large resistance of the gauges, which is roughly one or two orders of magnitude (~1kΩ) larger than that of metallic-layer piezoresistors (~10Ω). Taking into account the Johnson noise only, the SNR is given by

$$SNR_J = \frac{V_{out}}{\sqrt{S_J}} \propto \frac{\mathcal{V}_b \varepsilon}{\sqrt{4k_b TR}} \tag{7}$$

where *T* and *R* are the temperature and the gauge resistance respectively, $k_b$ is the Boltzmann constant, $V_b$ is the RMS value of the bias voltage. $V_{out}$ is proportional to $\delta R/R$ according to Eqn. (3). *SNR* for semiconducting gauge over *SNR* of metallic gauge can be simply expressed by,

$$\frac{SNR_{JS}}{SNR_{JM}} = \frac{\gamma_S V_{bS}}{\gamma_M V_{bM}} \sqrt{\frac{R_M}{R_S}} \tag{8}$$

Indexes $_S$ and $_M$ are for semiconductor gauge and metallic layer respectively.

At constant temperature considering the aforementioned resistances $V_{bS}$ can be 100 times larger than $V_{bM}$ because of the respective fusion temperature of silicon and metals. The $SNR_{JS}$ is then 10 times larger than $SNR_{JM}$. The gauge factor of silicon nanowires are much higher than the metallic layer gauges used as piezoresistive detection scheme for NEMS. The signal improvement is then much higher than the noise enhancement and the Johnson noise impact is limited.

*3.3. Allan deviation*

Usually NEMS is embedded in a phase locked loop (PLL) or a self-excited loop in order to monitor time evolution of their resonant frequency. The frequency stability of the overall system (e.g. of the NEMS and the supporting electronics) is characterized by the Allan deviation, defined as [9]

$$\delta\omega_0/\omega_0 = \sqrt{\frac{1}{(N-1)} \sum_1^N \left(\frac{\overline{\omega}_{i+1} - \overline{\omega}_i}{\omega_0}\right)^2} \tag{9}$$

where $\overline{\omega}_i$ is the average angular frequency in the i$^{th}$ time interval $\tau$, *N* is the number of independent frequency measurements, which is assumed to be a sufficiently large number. The mass resolution $\delta m$ is then $\sqrt{2}M_{eff}\,\delta\omega_0/\omega_0$ for 1s-integration time. The theoretical Allan deviation can be expressed as [6],

$$\left(\delta\omega_0/\omega_0\right)_{th} = 10^{-DR/20}/\sqrt{2}Q \tag{10}$$

For the experimental dynamic range, (DR) of 100 dB (see Fig.5) the ultimate Allan deviation would be around $1.5 \times 10^{-9}$ over 1s-integration time. For an effective mass of 140 fg (see TAB. 1) and a Q-factor of 5000, this would result in a potential mass resolution of $\delta m = \frac{M_{eff}}{Q} \cdot 10^{-(DR/20)} \approx 0.3$ zg at room temperature and at relatively low frequency (20 MHz).

The experimental Allan deviation was measured in open loop recording the phase variation of the electrical signal at the NEMS output. NEMS was driven at its resonant frequency (20MHz). Allan deviation was measured in three steps (for short, intermediate and long times). For low time constants (<0.1s), the integration time of the lock-in and the global acquisition time were 100 µs and 10s respectively. For larger time constants, they were set to 100 µs and 4000s (50000s) respectively. These adjustments remove the effect of the lock-in filtering that would artificially decrease the Allan deviation and ensure at least 100 points for each interval. We can also note that the smallest interval is set by the transient time $Q/f$ (i.e. ~250 µs in our case). Typical experimental data are shown in FIG 6. For mass sensing the study has to be focused on short times lower than 1s. Typically, we achieved an Allan deviation of $10^{-6}$ for $\tau$ =1s at room temperature. For long time constant, the minimum Allan deviation reaches $6.10^{-7}$. This value is quite a classical one reported in many papers (see [3], [21] for example) and might be considered as the experimental limit.

The large difference of three orders of magnitude between the expected value and the experimental Allan variance has to do with the fact that actuation is not present during thermomechanical noise measurement. The DR measurements therefore do not take into account noise contributions from the actuation voltage and the thermal bath. Considering both a typical silicon NEMS temperature coefficient of 50 ppm/K and an Allan deviation close to $10^{-6}$, the related thermal bath temperature fluctuations will be around $10^{-2}$ K. The effect of temperature fluctuations on cantilever measurements is well explained in Ref [22]. To get better frequency stability we think that the temperature fluctuation should be controlled at least below this value. It is also essential to suppress the background level as much as possible in order to reduce the additional phase noise that results from background fluctuations associated with electronic and temperature instabilities. The discrepancy between the Allan deviation obtained with eqn. (10) and the experimental data is an open question that is currently being studied.

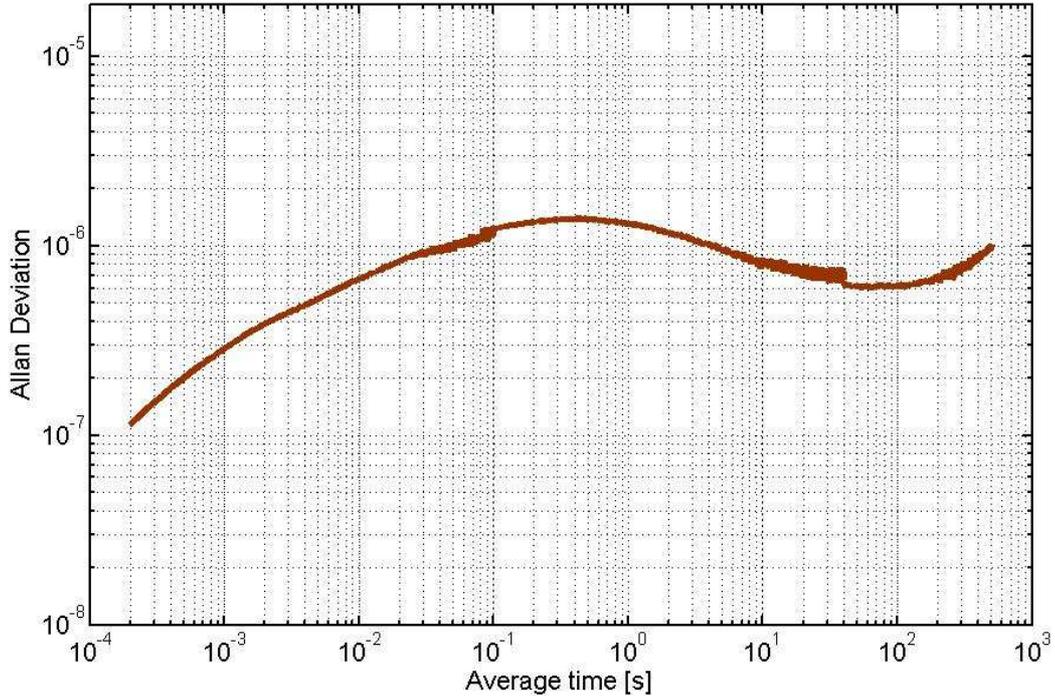

**FIG. 6** – Allan deviation measured in open loop for $V_{drive}$=1.5V and $V_{bias}$=1.5V.

### 4. Conclusion

In this paper, we demonstrate a new kind of detection scheme based on doped silicon nanowire strain gauges that are fully compatible with CMOS processes. This allows very large scale integration of devices in a straightforward manner. Measurements obtained with this approach are showing promising performances in term of frequency stability, dynamic range, and achievable mass resolution. The devices tested in this work were developed as prototypes and were not optimized for mass detection at this stage. Such NEMS have thus a great potential for future performance improvements and new applications opportunities. Further device optimization for lower mass and higher frequency, based on advanced top-down nanowire fabrication techniques [23] with expected giant gauge factors will lead to a resolution in the range of few zeptograms or less.

Several papers [3] [7] [10] have argued the importance of reducing the fundamental sources of noise by optimising the NEMS design. However, a tremendous effort is also needed to study and understand the coupling between NEMS and their environment (temperature fluctuation, packaging), which apparently limits the resolution so far.

This device with the lever arm architecture, symmetric piezoresistive gauges and decoupling between electrostatic actuation and piezoresistive detection makes the measurements more efficient and signal over noise ratio higher. Compared to metallic gauges, doped silicon gauges produce a much larger signal thanks to a much higher intrinsic gauge factor and larger allowed bias voltages (due to their higher resistance). The signal is thus much easier to detect while noise floor remains very low as it is dominated by thermomechanical and electronic, rather than Johnson, noise. Flicker noise ($1/f$ noise), which is often cited as a huge barrier for doped-silicon-based piezoresistive detection, is not an issue for RF resonance frequencies.

Very Large Scale Integration (VLSI) of devices described in this letter will enable a wide range of new devices, such as arrays of massively parallel oscillating NEMS, sensitive multigas sensors, and NEMS mass spectrometry with very low frequency dispersion less than 1%.